\documentclass[aps,prb,twocolumn,showpacs,preprintnumbers,amsmath,amssymb,superscriptaddress]{revtex4-1}

\usepackage{graphicx}
\usepackage{dcolumn}
\usepackage{bm}

\newcommand{\dyyrusix}{Dy$_{x}$Y$_{1-x}$Ru$_{2}$Si$_{2}$}
\newcommand{\dyyrusin}[2]{Dy$_{#1}$Y$_{#2}$Ru$_{2}$Si$_{2}$}
\newcommand{\tbyrusix}{Tb$_{x}$Y$_{1-x}$Ru$_{2}$Si$_{2}$}

\newcommand{\gdyrusix}{Gd$_{x}$Y$_{1-x}$Ru$_{2}$Si$_{2}$}

\newcommand{\femntiox}{Fe$_{x}$Mn$_{1-x}$TiO$_{3}$}

\newcommand{\kF}{k_{\mbox{\scriptsize F}}}
\newcommand{\Tg}{T_{\mbox{\scriptsize g}}}

\newcommand{\Tpeak}{T_{\mbox{\scriptsize peak}}}

\newcommand{\chieq}{\chi_{\mbox{\scriptsize eq}}}

\newcommand{\chinldc}{\chi_{\mbox{\scriptsize nl}}^{\mbox{\scriptsize dc}}}
\newcommand{\chindc}[1]{\chi_{#1}^{\mbox{\scriptsize dc}}}
\newcommand{\chiac}{\chi_{\mbox{\scriptsize ac}}}
\newcommand{\chinlac}{\chi_{\mbox{\scriptsize nl}}^{\mbox{\scriptsize ac}}}
\newcommand{\chinac}[1]{\chi_{#1}^{\mbox{\scriptsize ac}}}
\newcommand{\Rechiac}{\chi'_{\mbox{\scriptsize ac}}}
\newcommand{\Rechinlac}{\chi'_{\mbox{\scriptsize nl}}}
\newcommand{\Rechinac}[1]{\chi'_{#1}}

\newcommand{\Mdc}{M_{\mbox{\scriptsize dc}}}
\newcommand{\Mac}{M_{\mbox{\scriptsize ac}}}

\newcommand{\gammaMF}{\gamma_{\mbox{\scriptsize mf}}}
\newcommand{\betaMF}{\beta_{\mbox{\scriptsize mf}}}
\newcommand{\deltaMF}{\delta_{\mbox{\scriptsize mf}}}

\newcommand{\znuMF}{(z\nu)_{\mbox{\scriptsize mf}}}

\newcommand{\rhoMF}{\rho_{\mbox{\scriptsize mf}}}
\newcommand{\rhoSR}{\rho_{\mbox{\scriptsize sr}}}
\newcommand{\etaSR}{\eta_{\mbox{\scriptsize sr}}}

\begin{document}

\preprint{APS/123-QED}

\title{Dynamic scaling analysis of the long-range RKKY Ising spin glass \dyyrusix 
}

\author{Y. Tabata}%
\email{tabata.yoshikazu.7e@kyoto-u.ac.jp}%
\author{T. Waki}
\author{H. Nakamura}
\affiliation{Department of Materials Science and Engineering, Kyoto University, Kyoto 606-8501, Japan}

\date{\today}

\begin{abstract}
Dynamic scaling analyses of linear and nonlinear ac susceptibilities in a model magnet of the long-rang Ruderman-Kittel-Kasuya-Yoshida (RKKY) Ising spin glass (SG) \dyyrusin{0.103}{0.897} were examined. 
The obtained set of critical exponents, $\gamma$ $\sim$ 1, $\beta$ $\sim$ 1, $\delta$ $\sim$ 2, and $z\nu$ $\sim$ 3.4, indicates the SG phase transition belongs to a universality class different from that of either the canonical (Heisenberg) or short-range Ising SGs. The analyses also reveal a finite-temperature SG transition with the same critical exponents under a magnetic field and the phase-transition line $\Tg(H)$ described by $\Tg(H)$ $=$ $\Tg(0)(1-AH^{2/\phi})$ with $\phi$ $\sim$ 2. The crossover exponent $\phi$ obeys the scaling relation $\phi$ $=$ $\gamma + \beta$ within the margin of errors. These results strongly suggest spontaneous replica-symmetry breaking (RSB) with a {\it non- or marginal-mean-field universality class} in the long-range RKKY Ising SG. 
\end{abstract}

\pacs{75.50.Lk, 75.10.Nr, 75.40.Cx, 75.40.Gb}
\maketitle

\section{Introduction}
\label{intro}

The stability of a spin glass (SG) under a magnetic field is an important issue because it relates to the nature of the SG state, that is, {\it complex} or {\it simple}. The mean-field theory based on the infinite-range interaction model, the so-called Sherington-Kirkpatrick (SK) model \cite{SKmodel}, predicts a spontaneous {\it replica-symmetry breaking} (RSB) at the SG transition temperature $\Tg$ \cite{ParisiRSB}. In the RSB picture, the SG state is characterized by a complex free-energy landscape with multivalley structure and numerous thermodynamic equilibrium states, which are not related to each other by trivial symmetry operations \cite{MezardRSB}. On the other hand, a different SG picture was predicted on the basis of the phenomenological droplet theory \cite{FisherHuseDropletFirst}, where the replica symmetry (RS) is maintained and only two thermodynamic equilibrium states exist, which are trivially related to each other by the time-reversal operation. The former RSB SG is stable even under a finite magnetic field \cite{AT}, whereas, the latter RS SG state becomes unstable by an infinitely small magnetic field \cite{FisherHuseDropletFirst}. Hence, the stability of the SG under a magnetic field is a good {\it touchstone} to distinguish the RSB and RS SGs. 


In spite of many efforts and its scientific importance, the stability of the SG state under a magnetic field is still an open question, except for limited number models, such as the SK model.  The systems below the upper critical dimension ($<$ $d_{\mbox{\scriptsize u}}$ $=$ 6), including a three-dimensional (3D) system, are especially interesting because they can be compared with realistic systems directly. However, there is still no definitive consensus on this issue in such ``low"-dimensional systems. Indeed, most of the numerical studies did not show the SG transition in a field \cite{SasakiField,TJoergField,HGKatzgraberField,DLarsonField}, although several did \cite{LeuzziSGunderH,RABanosSGunderH,MBaityJesiSGunderH}.  The conflict may be due to quite large corrections to finite-size scaling. 

Experimentally, the stability of the SG under a magnetic field has also been examined for various type of systems \cite{NordbladFeMnTiO,JonssonFeMnTiO,NakatsukaFePhosphate,TabataDYRSJPSJ,TabataRYRS,CampbellTorque}: Ising-like, XY-like, and Heisenberg-like systems with a short-range (super)exchange interaction and long-range dipolar and Ruderman-Kittel-Kasuya-Yoshida (RKKY) interactions; however, the results are not conclusive yet. The difficulty of experimentally verifying the SG transition in a field is mostly due to the nonequilibrium effects due to their unavoidable long relaxation time. 

Here we introduce two former experimental studies using the same protocol to test the SG transition in a field by dynamic critical scaling analysis: one is an experiment for a model magnet of the short-range Ising SG \femntiox \cite{JonssonFeMnTiO} and the other is for that of the long-range RKKY Ising SG \dyyrusix \cite{TabataDYRSJPSJ}. According to the dynamic scaling hypothesis \cite{HohenDynamicCriticalPhenomena},  the characteristic relaxation time $\tau$ diverges when approaching the transition temperature $\Tg$ in a power-law fashion with the critical exponent $z\nu$. If the transition presents even in a finite magnetic field, the characteristic relaxation time at certain temperature $T$ and field $H$, $\tau(T, H)$, obeys 
\begin{equation}
\tau(T,H) = \tau_{0} \left[ T/\Tg(H) - 1 \right]^{-z\nu}, 
\label{DynamicCriticalScaling}
\end{equation}
with a field-dependent transition temperature $\Tg(H)$ and field-insensitive microscopic time-scale $\tau_{0}$ and $z\nu$. For both \femntiox\ and \dyyrusix , $\tau(T,H)$ was estimated from the frequency $\omega$ dependence of the ac susceptibility using the same protocol: $\chi'(T, H; 1/\tau)$ $=$ $(1-\alpha) \chieq(T,H)$, where $\chi'(T, H; \omega)$ is the real-part of the ac susceptibility and $\chieq(T,H)$ is the equilibrium susceptibility estimated from the dc-magnetization measurements. In \femntiox , $\tau$ obeys the dynamic scaling law (\ref{DynamicCriticalScaling}) only at zero field and, rather, obeys the scaling law based on the droplet theory where $\tau$ diverges only at a zero field below $\Tg(0)$ or at zero temperature in finite fields. On the other hand, $\tau(T,H)$ obeys Eq. (\ref{DynamicCriticalScaling}) in both zero and finite fields in \dyyrusix . These results strongly suggest that the short-range and long-range Ising SG belong to different universality classes, and the replica symmetry is broken only in the long-range one. 

In this paper, we report on an extended study to verify the SG transition under a magnetic field in \dyyrusin{0.103}{0.897} . The temperature region used for the scaling analysis in the previous study \cite{TabataDYRSJPSJ}, $\varepsilon$ ($\equiv$ $T/\Tg-1$) $>$ 0.2, was rather far from $\Tg$, and thus, the observed critical behavior might not be genuine. In this study, dynamic scaling analyses of the linear and nonlinear ac susceptibilities were examined by using data in the temperature $T$ regions closer to $\Tg$, where the scaling analysis works more appropriately. The analyses reveal the finite-temperature SG transition both in zero and finite magnetic fields, as well as in the previous study. The obtained set of the critical exponents is $\gamma$ $\sim$ 1, $\beta$ $\sim$ 1, $\delta$ $\sim$ 2, and $z\nu$ $\sim$ 3.4.  The crossover exponent $\phi$ $\sim$ 2, obtained from the field dependence of $\Tg(0)-\Tg(H)$ $\propto$ $H^{2/\phi}$, obeys the scaling law \cite{FisherSompolinskyScaling} $\phi$ $=$ $\gamma + \beta$ within a margin of error. The existence of the SG transition in a finite magnetic field indicates the spontaneous RSB in the long-range RKKY Ising SG. Nevertheless, the fulfillment of the scaling law $\phi$ $=$ $\gamma + \beta$ and the $z\nu$ slightly larger than the mean-field value \cite{SGreview} $\znuMF$ $=$ 2 suggest that the SG transition belongs to a different universality class with the mean-field one. 

\section{Experiments}
\label{exp}

A single crystalline sample of \dyyrusin{0.103}{0.897}\ grown using the Czochralski method with a tetra-arc furnace, which had been used for a previous study \cite{TabataDYRSJPSJ}, was reused for the present experiments. The ac and dc magnetization measurements were performed using the superconducting quantum interference device magnetometer (Quantum Design) equipped in the Research Center for Low Temperature and Material Science in Kyoto University. Both measurements were performed in the temperature and field region: 1.9 K $\leq$ $T$ $\leq$ 4.0 K and 0 Oe $\leq$ $H$ $\leq$ 800 Oe, respectively. The dc magnetization was measured 10 min after stabilization at certain temperature and field to obtain a thermodynamic equilibrium magnetization.
The ac susceptibility was measured with an ac field of 3 Oe and a frequency of 0.01 Hz $\leq$ $\omega$ $\leq$ 20 Hz.  The waiting time after the stabilization of temperature and field for the ac measurement was also 10 min. We confirmed that the ac susceptibility does not show waiting time dependence after 5 min within a margin of error down to 1.9 K, which is the lowest measurement temperature of this study. This result indicates that we can avoid aging effects, which are often observed in spin glass states, and we can capture the equilibrium dynamics in the present measurements. 

\section{Results and Analyses}
\label{res-ana}

\subsection{In zero magnetic field}
\label{inzerofield}

\begin{figure}
 \begin{center}
  \includegraphics[clip,width=8cm]{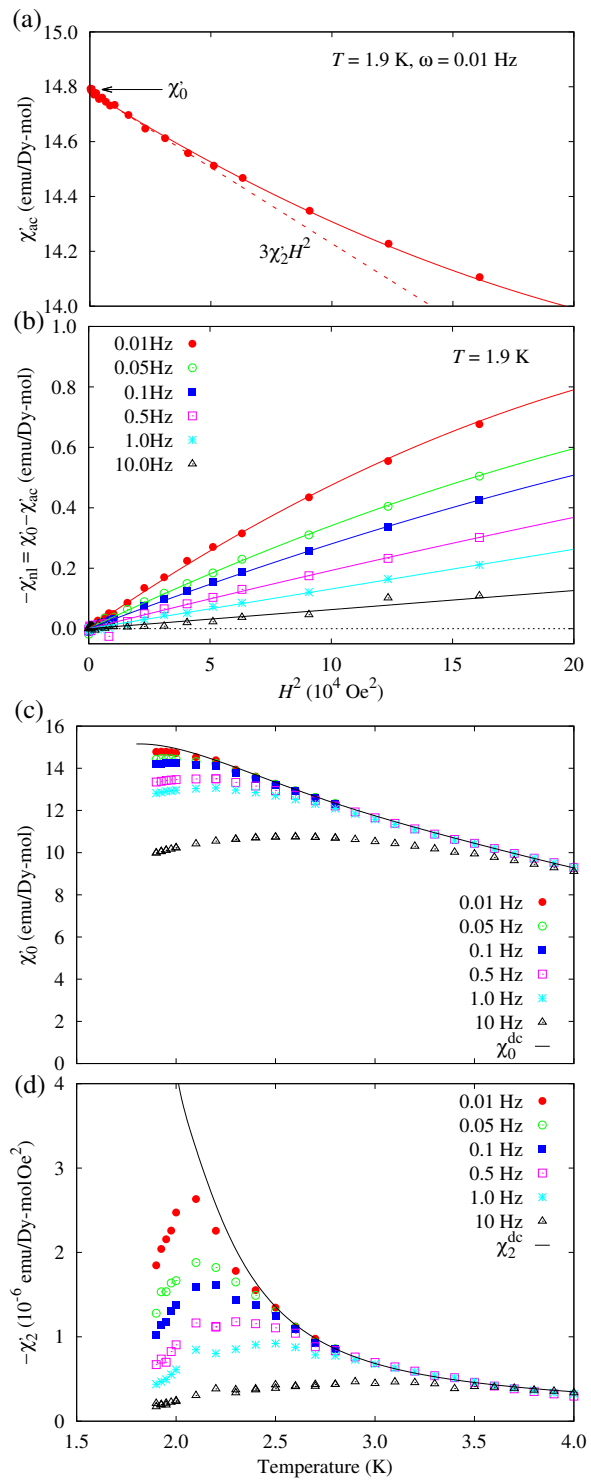}
 \caption{\label{LandNLAcSus}
 (a) The real part of the ac susceptibility $\Rechiac$ with $\omega$ $=$ 0.01 Hz at $T$ $=$ 1.9 K. (b) The real part of the full nonlinear ac susceptibility $\Rechinlac$ with several representative frequencies at $T$ $=$ 1.9 K. Solid lines in (a) and (b) represent the fitting results using Eq. (\ref{AcSusExp}) up to $n$ $=$ 2. Dashed line in (a) represents a leading term in Eq. (\ref{AcSusExp}), $3\Rechinac{2}H^{2}$. Temperature dependences of (c) the real part of the linear ac susceptibility $\Rechinac{0}$ and (d) first nonlinear ac susceptibility coefficient $\Rechinac{2}$ with several representative frequencies. Solid lines represent the  corresponding dc susceptibilities, $\chindc{0}$ and $\chindc{2}$.}
 \end{center}
\end{figure}

\begin{figure}
 \begin{center}
  \includegraphics[clip,width=7.5cm]{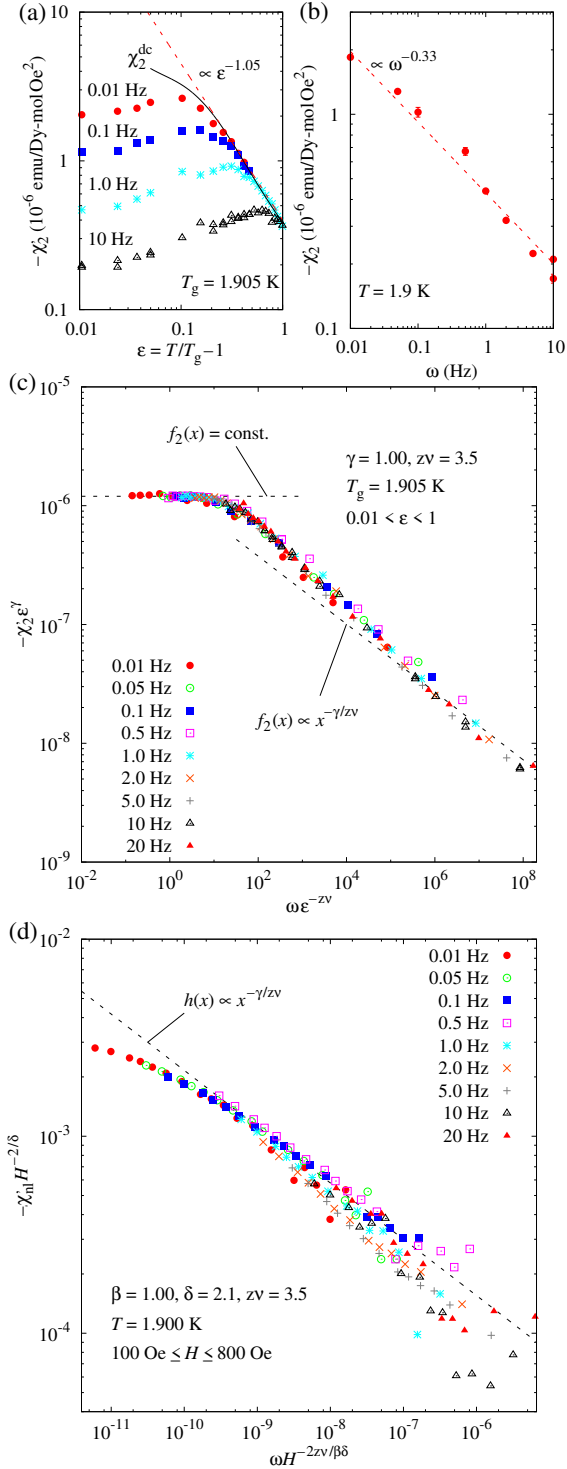}
 \caption{\label{DScalingZeroField}
 (a) Double-logarithmic plot of $\Rechinac{2}$ with several representative frequencies against the reduced temperature $\varepsilon$ $=$ $T/\Tg -1$ assuming $\Tg$ $=$ 1.905 K. The dashed line represents a divergent behavior of $\Rechinac{2}$ $\propto$ $\varepsilon^{-1.05}$. (b) Double-logarithmic plot of $\Rechinac{2}$ at $T$ $=$ 1.9 K against the frequency. The dashed line represents a divergent behavior of $\Rechinac{2}$ $\propto$ $\omega^{-0.33}$. (c) Dynamic scaling plot of $\Rechinac{2}$ in the form of $-\Rechinac{2}\varepsilon^{\gamma}$ vs $\omega\varepsilon^{-z\nu}$, described in Eq. (\ref{AcNLSDynamicScaling}). The dashed lines represent the asymptotes of the scaling function given in Eq. (\ref{AcNLSDynamicScalingFunc}). (d) Dynamics scaling plot of $\Rechinlac$ at $T$ $=$ 1.9 K in the form of $-\Rechinlac H^{-2/\delta}$ vs $\omega H^{-2z\nu/\beta\delta}$, described in Eq. (\ref{AcFNLSAtTgDynamicScaling}). The dashed line represents the asymptote of the scaling function given by Eq. (\ref{AcFNLSAtTgDynamicScalingFunc}).}
 \end{center}
\end{figure}

Here we show the experimental results and dynamic scaling analyses of nonlinear ac susceptibility and the imaginary part of the ac susceptibility to investigate the thermodynamic SG phase transition of \dyyrusin{0.103}{0.897} in zero magnetic field. In the zero-field limit, nonlinear susceptibility, corresponds to the SG susceptibility and the critical phenomena, divergences of the SG susceptibility with a static critical exponent $\gamma$ and the characteristic relaxation time with a dynamic critical exponent $z\nu$ when approaching the SG transition temperature $\Tg$, can be captured simultaneously by measuring the nonlinear ac susceptibility. The divergence of the characteristic relaxation time can also be captured by measuring the imaginary part of the ac susceptibility, which is related to the dynamic spin autocorrelation function by the fluctuation-dissipation theorem. The details of the dynamic scaling of the nonlinear ac susceptibility and the imaginary part of the ac susceptibility are described in Appendixes \ref{AppFNLSDS} and \ref{AppIACSUSDS}.

Linear and nonlinear dc susceptibilities $\chindc{2n}(T)$ are obtained by fitting the dc magnetization divided by the field $\Mdc(T,H)/H$ as a series function of $H^{2}$ in the form of 
\begin{equation}
\Mdc(T,H)/H \, = \, \sum_{n \geq 0} \chindc{2n}(T) H^{2n} , 
\label{DcSusExp}
\end{equation}
and these dc susceptibilities generally can be considered thermodynamic equilibrium ones. 
Similarly, linear and nonlinear ac susceptibilities $\chinac{2n}(\omega;T)$ are obtained by fitting the ac susceptibility in the dc field $\chiac(\omega;T,H)$ as a series function of $H^{2}$. Since $\chiac(\omega; T, H)$, dynamic response with small ac field at $T$ and $H$, can be recognized as a differential magnetization with a frequency of $\omega$ $d\Mac(\omega; T, H)/dH$ \cite{LevyAgMnPRB}, it is given by a series function of $H^{2}$ in the form of, 
\begin{align}
\chiac(\omega;T,H) &= \frac{d\Mac(\omega;T,H)}{dH} \notag \\
 &= \sum_{n \geq 0} (2n+1)\chinac{2n}(\omega;T)H^{2n} . 
\label{AcSusExp}
\end{align}
For instance, the field dependence of the real part of the ac susceptibility $\Rechiac$ with $\omega$ $=$ 0.01 Hz at $T$ $=$ 1.9 K is shown in Fig. \ref{LandNLAcSus}(a), which is well fitted by using Eq. (\ref{AcSusExp}) up to $n$ $=$ 2 denoted as a solid line. The linear ac susceptibility $\Rechinac{0}$ ($n$ $=$ 0) is obtained as the intercept of the vertical axis and the first nonlinear ac susceptibility coefficient $\Rechinac{2}$ ($n$ $=$ 1), the leading nonlinear term, is obtained as the initial slope of the $\Rechiac$-curve. The real part of the full nonlinear ac susceptibility $\Rechinlac(\omega;T,H)$ $\equiv$ $\Rechiac(\omega;T,H) - \Rechinac{0}(\omega;T)$, including all nonlinear terms in the ac susceptibility, with several frequencies at $T$ $=$ 1.9 K is shown in Fig. \ref{LandNLAcSus}(b) with the nonlinear part of the fitting results using Eq. (\ref{AcSusExp}). Strong suppression of $\Rechinlac(\omega;T,H)$ with increasing frequency is found.  As discussed later, the nonlinear ac susceptibility does not merge into the static one even with 0.01 Hz below 2.5 K, in the vicinity of $\Tg$, indicating the time scale of the critical spin dynamics in this temperature region is more than 100 s. 

Figures \ref{LandNLAcSus}(c) and \ref{LandNLAcSus}(d) show temperature dependences of the real parts of the linear susceptibility and the first nonlinear ac susceptibility coefficient, $\Rechinac{0}(\omega;T)$ and $\Rechinac{2}(\omega;T)$. Both susceptibilities exhibit frequency dependences below 4 K. Especially, strikingly strong  frequency dependence is found in $\Rechinac{2}(\omega;T)$, which corresponds to a dynamic susceptibility of the SG order parameter. Indeed, $\Rechinac{2}(\omega;T)$ with $\omega$ $=$ 0.01 Hz deviates from the nonlinear dc susceptibility coefficient $\chindc{2}(T)$ below 2.5 K, whereas $\Rechinac{0}(\omega;T)$ with 0.01 Hz almost collapses on $\chindc{0}(T)$. Hence, the observed slow dynamics is considered to be due to the critical slowing down of the SG transition. The temperature dependence of $\Rechinac{2}(\omega;T)$ is also much stronger than that of $\Rechinac{0}(\omega;T)$ and shows a peak at a frequency-dependent temperature $\Tpeak(\omega)$, which corresponds to a dynamic spin-freezing temperature where spins are frozen within the time scale of $1/\omega$ and is extrapolated to a thermodynamic equilibrium SG transition temperature $\Tg$ with $\omega$ $\rightarrow$ 0. The peak of $\Rechinac{2}(\omega; T)$ becomes sharper and its temperature begins to fall with decreasing frequency and seems to be extrapolated to a certain finite temperature $\sim$ 1.9 K. This suggests the thermodynamic SG transition at around $\sim$ 1.9 K at zero field.


Figure \ref{DScalingZeroField}(a) shows a double-logarithmic plot of the real part of the first nonlinear ac susceptibility coefficient $\Rechinac{2}$ against the reduced temperature $\varepsilon$ $=$ $T/\Tg -1$ with assuming $\Tg$ $=$ 1.905 K, which is actually obtained by the dynamic scaling analysis of $\Rechinac{2}$ described later. $\Rechinac{2}$ exhibits a $\varepsilon^{-1.05}$ behavior in the high-temperature region, denoted by a dashed line. Similar $\varepsilon^{-1}$-like behavior is observed whenever $\Tg$ is assumed to be near 1.9 K. A double-logarithmic plot of $\Rechinac{2}$ at 1.9 K against $\omega$ is shown in Fig. \ref{DScalingZeroField} (b). In the plot, $\Rechinac{2}$ diverges down to $\omega$ $\rightarrow$ 0 as $\omega^{-0.33}$. These divergent behaviors of $\Rechinac{2}$ against $\varepsilon$ and $\omega$ are considered its critical divergences in the asymptotic region for $T$ $\rightarrow$ $\Tg$ and $\omega$ $\rightarrow$ $0$, expressed in Eq. (\ref{AcNLSAsymptote}) in Appendix \ref{AppFNLSDS}, which suggest critical exponents $\gamma$ and $z\nu$ are roughly $\sim$ 1 and $\sim$ 3 respectively. We tried to analyze the experimental data of $\Rechinac{2}(\omega ; T)$ in the range of 0.01 $<$ $\varepsilon$ $<$ 1 and 0.01 Hz $\leq$ $\omega$ $\leq$ 20 Hz using a dynamic scaling plot in the form of $-\Rechinac{2} \varepsilon^{\gamma}$ vs $\omega\varepsilon^{-z\nu}$, described in Eq. (\ref{AcNLSDynamicScaling}). The best scaling plot is shown in Fig. \ref{DScalingZeroField}(c). The experimental data well collapsed on a unique curve reveals that the dynamic critical scaling of $\Rechinac{2}$ works very well when assuming a finite SG transition temperature. Furthermore, the experimental data obey the asymptotes of the scaling function given in Eq. (\ref{AcNLSDynamicScalingFunc}), as denoted by dashed lines in Fig. \ref{DScalingZeroField}(c), which corresponds to the divergent behaviors for $\omega$ $\rightarrow$ 0 and $\varepsilon$ $\rightarrow$ 0 shown in Figs. \ref{DScalingZeroField}(a) and \ref{DScalingZeroField}(b). The obtained transition temperature and critical exponents are $\Tg$ $=$ $1.905(7)$ K and $\gamma$ $=$ $1.00(5)$ and $z\nu$ $=$ $3.5(1)$, respectively.  

Figure \ref{DScalingZeroField}(d) shows a dynamic scaling plot of the real part of the full nonlinear ac susceptibility very close to $\Tg$ [$T$ $=$ 1.9 K shown in Fig. \ref{LandNLAcSus}(a)] in the form of $-\Rechinlac H^{-2/\delta}$ vs $\omega H^{-2z\nu/\beta\delta}$, given in Eq. (\ref{AcFNLSAtTgDynamicScaling}). In this plot, the experimental data in the region of $100$ Oe $\leq$ $H$ $\leq$ 800 Oe and 0.01 Hz $\leq$ $\omega$ $\leq$ 20 Hz collapses on a unique curve. The obtained critical exponents are $\delta$ $=$ $2.1(1)$ and $z\nu/\beta\delta$ $=$ $1.62(5)$. The dynamics of \dyyrusin{0.103}{0.897} near $\Tg$ is too slow to observe the constant behavior of the scaling function for $\omega$ $\rightarrow$ $0$, whereas the power-law behavior for $H$ $\rightarrow$ $0$ is clearly seen, as denoted by a dashed line in Fig. \ref{DScalingZeroField}(d). The exponent $\beta$ is derived to be $1.00(5)$ from the above-mentioned dynamic scaling analyses of $\chi'_{2}(\omega; T)$ and $\Rechinlac(\omega;\Tg,H)$. The obtained set of static critical exponents, $\gamma$, $\beta$, and $\delta$, obeys the scaling law $\gamma + \beta$ = $\beta\delta$ within the margin of error [$\gamma + \beta$ $=$ $2.00(7)$ and $\beta\delta$ $=$ $2.10(12)$]. The results, shown in Figs. \ref{DScalingZeroField}(c) and \ref{DScalingZeroField}(d), indicate that the dynamic scaling of the experimentally obtained nonlinear ac susceptibility works very well and confirm the thermodynamic SG phase transition at zero field. 


\begin{figure}[ht]
 \begin{center}
  \includegraphics[clip,width=8cm]{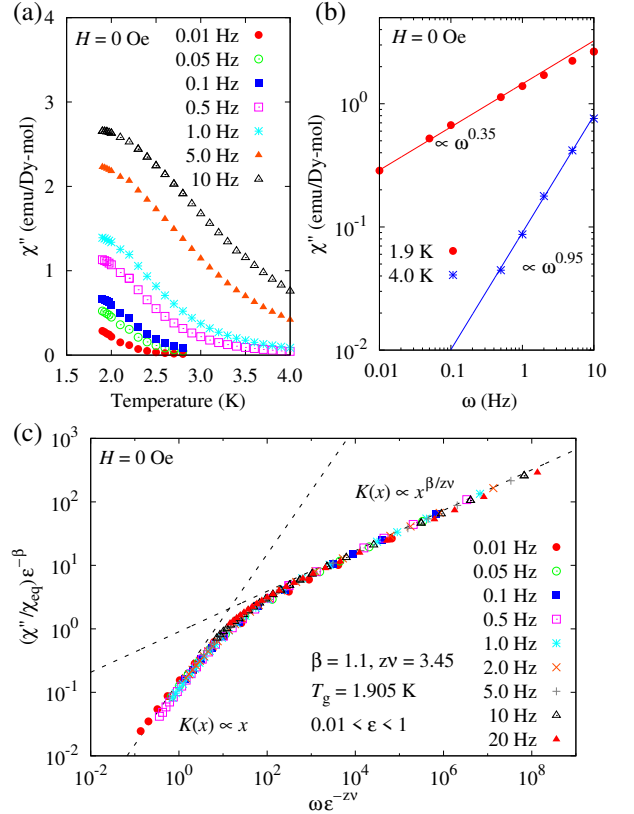}
 \caption{\label{ImAcSus}
 (a) Temperature dependence of the imaginary part of the ac susceptibility $\chi''$ with several representative frequencies at a zero field. (b) Double-logarithmic plot of $\chi''$ at $T$ $=$ 1.9 K and 4.0 K against the frequency. Solid lines represent nonanalytic and nearly analytic behaviors at 1.9 and 4.0  K, $\omega^{0.35}$ and $\omega^{0.95}$, respectively. (c) Dynamic scaling plot of $\chi''$ in the form of $(\chi''/\chieq)\varepsilon^{-\beta}$ vs $\omega\varepsilon^{-z\nu}$, described in Eq. (\ref{ImSusDynamicScalingEpsilon}). Dashed lines represent the asymptotes of the scaling function given in Eq. (\ref{ImSusDynamicScalingFuncAsympt}). }
 \end{center}
\end{figure}


The temperature dependence of the imaginary part of the ac susceptibility at zero field $\chi''(\omega; T)$ with various frequencies is shown in Fig. \ref{ImAcSus} (a). Corresponding to the appearance of the frequency dependence of the real part of the ac susceptibility shown in Fig. \ref{LandNLAcSus} (b), $\chi''(\omega;T)$ is emergent below 4 K. The $\omega$ dependences of $\chi''(\omega;T)$ at 1.9 and 4.0 K are shown in Fig \ref{ImAcSus}(b), exhibiting $\omega^{0.35}$- and $\omega^{0.95}$-behaviors at 1.9 K ($\approx$ $\Tg$) and 4.0 K ($\approx$ $2\Tg$), respectively. The strongly nonanalytic behavior close to $\Tg$ is a manifestation of the critical dynamics described in Eq. (\ref{ImSusAsympt}) in Appendix \ref{AppIACSUSDS}, suggesting $\beta/z\nu$ $=$ 0.35. The dynamic scaling plot of $\chi''(\omega; T)$ in the form of $(\chi''/\chieq)\varepsilon^{-\beta}$ vs $\omega\varepsilon^{-z\nu}$ described in Eq. (\ref{ImSusDynamicScaling}) is shown in Fig. \ref{ImAcSus} (c). In this analysis, the transition temperature $\Tg$ is fixed to be 1.905 K obtained from the dynamic scaling of the first nonlinear ac susceptibility coefficient. The scaling with $\beta$ $=$ $1.1(1)$ and $z\nu$ $=$ $3.45(8)$ reveals that the experimental data in $0.01$ $<$ $\varepsilon$ $<$ $1.0$ and $0.01$ Hz $\leq$ $\omega$ $\leq$ 20 Hz collapse on a unique curve very well. The asymptotic behaviors of the scaling function $K(x)$ expressed in Eq. (\ref{ImSusDynamicScalingFuncAsympt}) are also clearly found, which are represented by dashed lines in Fig. \ref{ImAcSus} (c). The obtained critical exponents are consistent with those obtained in the analyses of $\Rechinac{2}$ and $\Rechinlac$ within a margin of error. The result of this scaling analysis of $\chi''(\omega;T)$ affirms the zero field SG transition at $\Tg$ $=$ 1.905 K. 

\subsection{In finite magnetic fields}
\label{infinitefields}

\begin{figure}
 \begin{center}
  \includegraphics[clip,width=8cm]{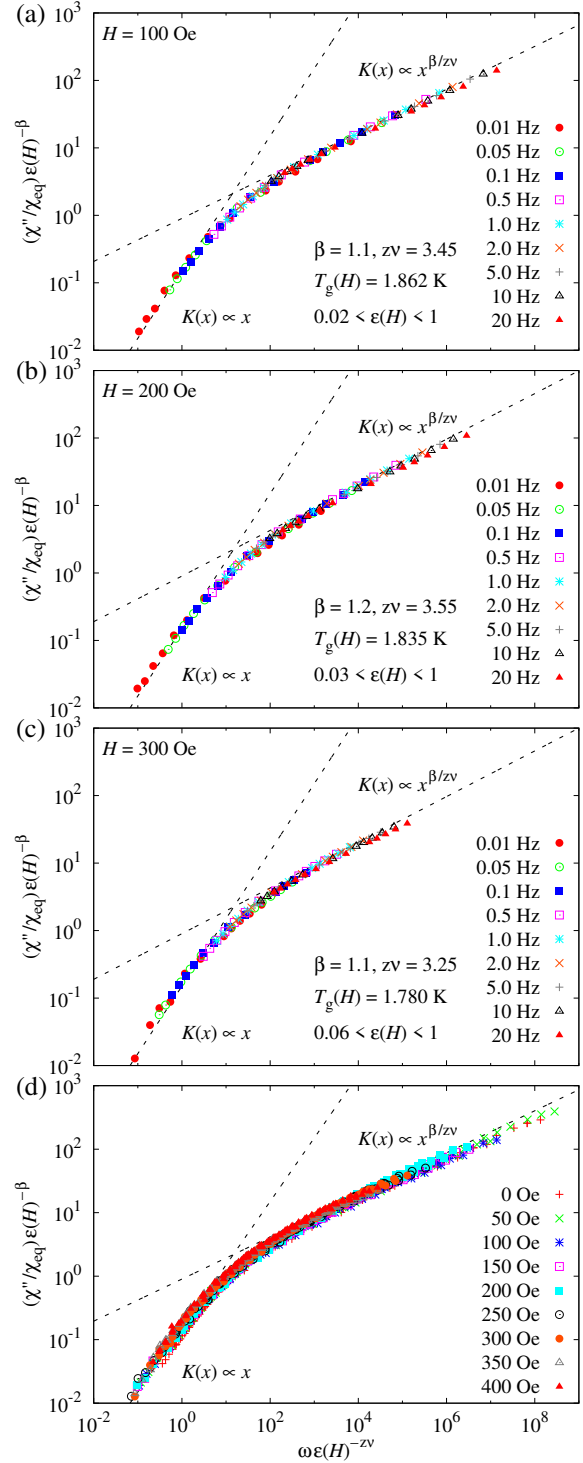}
 \caption{\label{ImAcSusField}
 Dynamic scaling plots of $\chi''$ in the form of $(\chi''/\chieq)\varepsilon(H)^{-\beta}$ vs $\omega\varepsilon(H)^{-z\nu}$ at $H$ $=$ (a) 100, (b) 200, and (c) 300 Oe. Dashed lines represent the asymptotes of the scaling function given in Eq. (\ref{ImSusDynamicScalingFuncAsympt}). (d) Scaling plot of $\chi''$ for all fields is simultaneously shown to see a field-independent feature of the scaling function $K(x)$. Dashed lines in (d) are the asymptotes using the mean value $\overline{\beta}$ $=$ 1.11 and $\overline{z\nu}$ $=$ 3.35 shown in Fig. \ref{HTpdiagram} (c).}
 \end{center}
\end{figure}

\begin{figure}[ht]
 \begin{center}
  \includegraphics[clip,width=8cm]{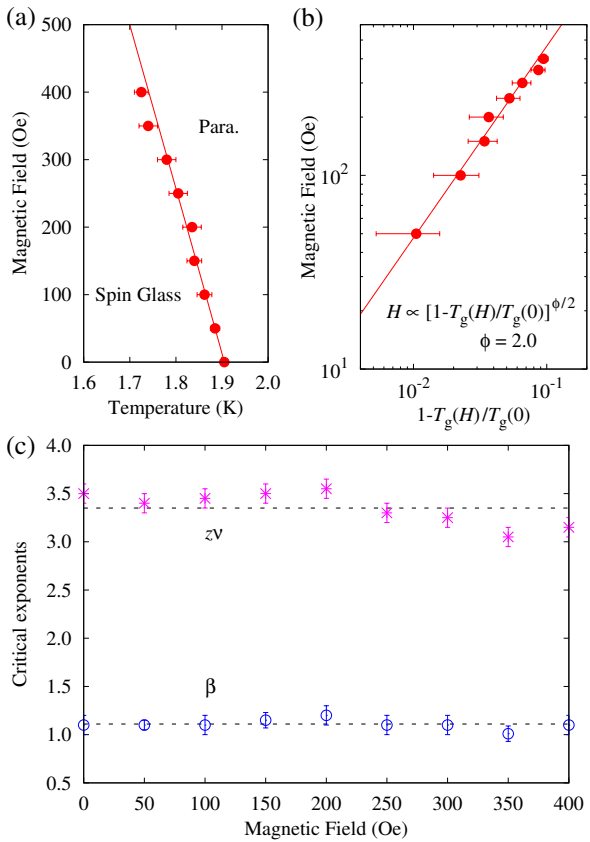}
 \caption{\label{HTpdiagram}
 (a) $HT$ phase diagram of \dyyrusin{0.103}{0.897} derived from the dynamic scaling analyses of $\chi''$.  (b) Double-logarithmic plot of $\Tg(H)$ in the form of $H$ vs $1-\Tg(H)/\Tg(0)$. Solid lines in (a) and (b) represent the fitting result using Eq. (\ref{FuncTgH}). (c) Field dependences of the critical exponents $\beta$ and $z\nu$. Dashed lines represent the mean value $\overline{\beta}$ $=$ 1.11 and $\overline{z\nu}$ $=$ 3.35.}
 \end{center}
\end{figure}

Next, we try carrying out the dynamic scaling of $\chi''(\omega;T,H)$ in finite fields. If the thermodynamic SG transition exists in finite fields and the characteristic relaxation time $\tau(T,H)$ obeys the critical divergence expressed in Eq. (\ref{DynamicCriticalScaling}),  the dynamic scaling of $\chi''(\omega;T,H)$ works with finite transition temperatures $\Tg(H)$ as well as in zero field. It should be noted that the critical divergence of the nonlinear susceptibility coefficient is suppressed in the presence of magnetic field, and thus, observation of the divergence of $\tau(T,H)$ through the dynamic scaling of $\chi''(\omega;T,H)$ is the best known way to capture the SG phase transition in a finite field.

Figures \ref{ImAcSusField}(a) - \ref{ImAcSusField}(c) show dynamic scaling plots of $\chi''(\omega;T,H)$ at the representative fields, $H$ $=$ 100, 200, 300 Oe. The SG transition temperature $\Tg(H)$ and critical exponents $\beta$ and $z\nu$ are evaluated by these dynamic scaling analyses separately for each field. The experimental data collapse on a unique curve at each field with the asymptotic behaviors $\propto$ $x$ for $x$ $\rightarrow$ 0 and $\propto$ $x^{\beta/z\nu}$ for $x$ $\rightarrow$ $\infty$. The best scalings shown in the figures are obtained with almost identical sets of critical exponents, $\beta$ $\sim$ 1.1 and $z\nu$ $\sim$ $3.4$, and monotonically reduced $\Tg(H)$ with increasing field. The field dependences of the obtained critical exponents and SG transition temperatures are shown in Figs. \ref{HTpdiagram}(a) and \ref{HTpdiagram}(c). The critical exponents are quite insensitive to the magnetic field. Not only the critical exponents but also the scaling function $K(x)$ are almost universal against the magnetic field, as shown in Fig. \ref{ImAcSusField}(d). These results demonstrate the existence of the thermodynamic SG phase transition in both zero and finite fields with the same universality class in \dyyrusin{0.103}{0.897}, indicating the spontaneous RSB. In the $HT$ phase diagram [Fig. \ref{HTpdiagram}(a)] an Almeida-Thouless-like $\Tg(H)$ is found, which is described by the functional form of 
\begin{align}
\Tg(H) &= \Tg(0)\left(1 - AH^{2/\phi}\right)
\label{FuncTgH}
\end{align}
as shown in Fig. \ref{HTpdiagram}(b). The crossover exponent $\phi$ is estimated as $\phi$ $=$ $2.0(2)$, which is discussed in detail in Sec. \ref{discuss}. 

 
\section{Discussion}
\label{discuss}

\subsection{Critical exponents and universality class}
\label{CriticalExponents}

Here we discuss the universality class of the RKKY Ising SG under zero and finite fields based on the static and dynamic critical exponents obtained in the present experiments. 

The obtained set of static critical exponents, $\gamma$, $\beta$, and $\delta$, is very similar to that of the mean-field SK model, $\gammaMF$ $=$ 1, $\betaMF$ $=$ 1, and $\deltaMF$ $=$ 2. The value of $\beta$ does not vary in the field, being within 1.0 $-$ 1.2.  Hence, \dyyrusin{0.103}{0.897}, the model magnet of the long-range RKKY Ising SG, seems to belong to the mean-field universality class, from the viewpoint of the static critical phenomena. On the other hand, the crossover exponent $\phi$, defined from the field dependence of the SG transition temperature as $H^{2}$ $\propto$ $\left[1 - \Tg(H)/\Tg(0)\right]^{\phi}$ derived from Eq. (\ref{FuncTgH}), is estimated to be 2.0(2), as shown in Fig. \ref{HTpdiagram}(b),  which coincides with $\gamma + \beta$ $\sim$ 2. According to the Fisher-Sompolinsky scaling \cite{FisherSompolinskyScaling}, the scaling relation $\phi$ $=$ $\gamma + \beta$ is satisfied only below the upper critical dimension $d_{\mbox{\scriptsize u}}$ $=$ 6, which is violated above $d_{\mbox{\scriptsize u}}$ and continuously connected to the SK value \cite{AT} for $d$ $\rightarrow$ 8 : $\phi$ $=$ $d/2-1$ for 6 $\leq$ $d$ $\leq$ 8 and $\phi$ $=$ 3 for $d$ $\geq$ 8. The validity of the scaling relation $\phi$ $=$ $\gamma + \beta$ in our results indicates that the RKKY Ising SG belongs to the non- or marginal-mean-field universality class.  The dynamic critical exponent $z\nu$ $\sim$ 3.4 being larger than the mean-field value $\znuMF$ $=$ 2 reinforces this argument. If the system belongs to a non-mean-field universality class, under $d_{\mbox{\scriptsize u}}$, the hyperscaling law is applicable. In our \dyyrusin{0.103}{0.897}, we can derive the critical exponent of the correlation length $\nu$ $=$ $1.07(5)$ and anomalous dimension $\eta$ $=$ $1.07(7)$ using the hyperscaling relations $\gamma$ $=$ $\nu(2-\eta)$ and $d\nu$ $=$ $2-\alpha$ $=$ $2\beta+\gamma$. Similar static critical exponents were obtained in other RKKY Ising SG materials such as \tbyrusix\ and \gdyrusix \cite{TabataRYRS}, indicating the universality of the RKKY Ising SG.

The universality class of \dyyrusin{0.103}{0.897} is different from the two archetypes of the real SG systems, the canonical SG and \femntiox . The canonical SG is a group of SG materials consisting of a nonmagnetic metal with low-concentration magnetic impurities, such as Au(Fe), Cu(Mn), Ag(Mn), and Pt(Mn), which is categorized into the RKKY Heisenberg SG with weak random anisotropy. Most of the canonical SG materials exhibit a universal critical behavior with critical exponents $\gamma$ $\sim$ 2, $\beta$ $\sim$ 1, $\delta$ $\sim$ 3, and $z\nu$ $\sim$ 7 \cite{BouchiatAgMnJPhys,LevyAgMnPRL,TaniguchiAuFeJPSJ,TaniguchiPtMnJPconf}. On the other hand, \femntiox\ is a model magnet of the Ising SG with short-range (superexchange) interaction and shows a SG transition with critical exponents $\gamma$ $\simeq$ 4.0, $\beta$ $\simeq$ 0.54, $\delta$ $\simeq$ 8.4, and $z\nu$ $\simeq$ 10 \cite{GunnarssonFeMnTiO,NordbladFeMnTiO,JonssonFeMnTiO}. These differences of the critical exponents are attributed to differences of ``{\it relevant parameters}" of the critical phenomena, such as spin dimension, spatial dimension, and spatial dependence of interaction. The spin dimension $N$ is different between \dyyrusin{0.103}{0.897} and the canonical SG, for which $N$ $=$ 1 (Ising) and 3 (Heisenberg) respectively, whereas the spatial dependence of interaction is the same: the RKKY interaction works in both. On the other hand, between \dyyrusin{0.103}{0.897} and \femntiox , the spin dimension is the same, whereas the spatial dependence of interaction is different. The critical exponents of these materials are listed in Table \ref{TabCE}. In Table \ref{TabCE}, $\nu$ and $\eta$ in the canonical SG and \femntiox , obtained using hyperscaling relations, are also shown. 

\begin{table*}[ht]
\begin{tabular}{c|c|c|c|c|c|c|c} 
Compound or model & $\gamma$ & $\beta$ & $\delta$ & $z\nu$ & $\nu$ & $\eta$  & Ref. \\ \hline
\dyyrusix\ (Ising, RKKY) & 1.00(5) & 1.1(1) & 2.1(1) & 3.45(8) & 1.07(5) & 1.07(7) & this work\\ 
\femntiox\ (Ising, SE) & 4.0 & 0.54 & 8.4 & $\sim$ 11 & 1.7 & $-0.35$ & \onlinecite{GunnarssonFeMnTiO,NordbladFeMnTiO,JonssonFeMnTiO} \\ 
canonical SGs (Heisenberg, RKKY) & 1.5 - 2.2 & 0.9 - 1.0 & 2.5 - 3.3 & $\sim 7$ & 1.2 - 1.4 & 0.4 - 0.7 & \onlinecite{LevyAgMnPRB,BouchiatAgMnJPhys,LevyAgMnPRL,TaniguchiAuFeJPSJ,TaniguchiPtMnJPconf}\\ \hline
simulation (Ising, SR) & 6.0 - 6.5 & 0.52 - 0.8 & 6.3 - 8.8 & $\sim 14$ & 2.35 - 2.72  & $-0.384$ - $-0.40$ & \onlinecite{CampbellExtendedScaling,HasenbuschIsingPRB,BaityJesiIsingPRB,BlundellIsingDSJPhysA} \\
simulation (Ising, dipolar) & $-$ & $-$ & $-$ & $-$ & 0.95 - 1.28 & $-$ & \onlinecite{TamDipolarIsing,AlonsoDipolarIsing,AndresenDipolarIsing} \\
simulation (Heisenberg, SR) & 2.6 & 0.95 & 3.7 & $-$ & 1.49 & 0.28 & \onlinecite{NakamuraHeisenbergPRE}\\ 
simulation (Heisenberg, RKKY) & 2.0 & 0.98 & 3.0 & $-$ & 1.3 & 0.5 & \onlinecite{ZhangRKKYHeisenberg} \\
simulation (Heisenberg, dipolar) & 0.72 & 1.4 & 1.5 & $-$ & 1.2 & 1.4 & \onlinecite{StasiakDipolarHeisenberg} \\ \hline
mean-field & 1 & 1 & 2 & 2 & 1/2 & 0 & \onlinecite{SGreview} \\ \hline
\end{tabular}
\caption{\label{TabCE}
Critical exponents obtained in various SG materials and theoretical models. The listed critical exponents in the present work are the values obtained from the analyses of $\Rechinac{2}$ ($\gamma$), $\Rechinlac$ ($\delta$), and $\chi''$ at zero field ($\beta$ and $z\nu$). Some exponents are derived via the scaling and hyperscaling relations by using reported exponents. SE and SR means superexchange and short-range interactions, respectively}
\end{table*}

Here, we should emphasize the difference between \dyyrusin{0.103}{0.897}\ and \femntiox , the long-range RKKY and short-range Ising SGs. Smaller $\nu$ and larger $\eta$ with the sign changed to positive are estimated in the long-range RKKY \dyyrusin{0.103}{0.897} . Especially, the large positive $\eta$ in the RKKY Ising SG causes the marginal mean-field-like critical phenomena. Some numerical works \cite{TamDipolarIsing,AlonsoDipolarIsing,AndresenDipolarIsing,StasiakDipolarHeisenberg,ZhangRKKYHeisenberg} also showed that the long-range interaction, such as RKKY and dipolar interactions, systems have a criticality different from that in corresponding short-range interaction systems. For instance, smaller $\nu$ is estimated in the dipolar Ising SG: $\nu$ $\sim$ 1 and 2.5 in the dipolar \cite{TamDipolarIsing,AlonsoDipolarIsing,AndresenDipolarIsing} and short-range Ising systems \cite{CampbellExtendedScaling,HasenbuschIsingPRB,BaityJesiIsingPRB}, respectively. In addition, slightly smaller $\nu$ and larger $\eta$ are estimated in the dipolar \cite{StasiakDipolarHeisenberg} and RKKY \cite{ZhangRKKYHeisenberg} Heisenberg systems compared with those in the short-range system \cite{NakamuraHeisenbergPRE,CommentOnHeisenbergSG}. The critical exponents obtained in these numerical works are also listed in Table \ref{TabCE}. Experimental and numerical studies on the criticality of long-range SGs in three dimensions are restrictive; for instance, $\eta$ has not been numerically estimated in the long-range Ising SGs, and only limited Ising compounds have been examined experimentally; thus, the comparison between them should be limited. Nevertheless, we should note that the same characteristic trends in long-range interaction systems, smaller $\nu$ and larger $\eta$, are found in both experimental and numerical works.   

On the other hand, the criticality of the RKKY Ising SG in three dimensions should belong to the same universality class as the corresponding short-range Ising SG, in accordance with scaling theory \cite{KotliarPRBOneDLRCP,BrayRG,LeuzziJPhysAOneDLRCP,LarsonPRBOneDLRCP,BanosPRBLRCP,BerganzaPRBLRCP}, which apparently contradicts our experimental findings. The system with the $\rho$th power decaying ($r^{-\rho}$) long-range interaction in $d$ dimensions can be broken into three regimes: (i) the mean-field regime $d$ $<$ $\rho$ $<$ $\rhoMF(d)$, where the mean-field criticality is correct, (ii) the long-range regime $\rhoMF(d)$ $<$ $\rho$ $<$ $\rhoSR(d)$, where a criticality different from the corresponding short-range one appears, and (iii) the short-range regime $\rhoSR(d)$ $<$ $\rho$, where the short-range criticality is found. In the SG system, the upper bound of the mean-field regime is given by $\rhoMF(d)$ $=$ $4d/3$ \cite{KotliarPRBOneDLRCP}. In the long-range regime, the correspondence between the $d$-dimensional long-range system and $D$-dimensional short-range system, $D$ $=$ $[2-\etaSR(D)]d/(\rho-d)$, is conjectured \cite{LarsonPRBOneDLRCP,BanosPRBLRCP,BerganzaPRBLRCP}, where $\etaSR(D)$ is the anomalous dimension in the short-range system. The upper bound of the long-range regime is given by the condition of $D$ $=$ $d$ as $\rhoSR(d)$ $=$ $d + 2 - \etaSR(d)$. The RKKY interaction decays as $J_{ij}$ $\sim$ $\cos(\kF r_{ij})/r_{ij}^{3}$, and its variance decays as $\overline{J_{ij}^{2}}$ $\propto$ $r_{ij}^{-6}$. Hence, the RKKY Ising SG system is a system with $d$ $=$ 3 and $\rho$ $=$ 6. This $\rho$ is apparently larger than $\rhoSR(3) = 5 - \etaSR(3)$ $=$ 5.35, where the anomalous dimension obtained in \femntiox\ shown in Table \ref{TabCE} is used, and $\rhoMF(3)$ $=$ 4. In a finite field, a different anomalous dimension $\etaSR^{h}$ and consequent different $\rhoSR^{h}$ can be expected, which can be lower than the zero-field $\rhoSR$ in any known case \cite{LeuzziPRBLRCPunderH}. As a consequence, the short-range criticality in the RKKY Ising SG in both zero and finite fields is theoretically expected. This significant discrepancy between our experimental findings and theoretical expectation for the criticality in the RKKY Ising SG is an open question and should be fixed in the future.

\subsection{Comparison with the previous work}
\label{PreviousWork}

Finally, we compare the present results with our previous work \cite{TabataDYRSJPSJ} where the mean-field-like criticality $\gamma$ $\sim$ 1, $\beta$ $\sim$ 1, $\delta$ $\sim$ 2, $z\nu$ $\sim$ 2, and $\phi$ $\sim$ 2.4 was reported. In the present work, we obtained almost the same static critical exponents, a larger dynamics critical exponent, and a smaller crossover exponent. Here we discuss possible reasons why slightly different criticality is found, and the present result is more genuine. 

The most significant difference between the two studies is that of the parameter ($T$, $H$, and $\omega$) regions of the dynamic scaling analyses. The analyses in the present study were conducted using experimental data closer to $\Tg$ and lower frequency, namely, $0.01$ $\lesssim$ $\varepsilon$ $\lesssim$ $1$ and $0.01$ Hz $\leq$ $\omega$ $\leq$ $20$ Hz in the present study and $0.1$ $\lesssim$ $\varepsilon$ $\lesssim$ $2$ and $0.1$ Hz $\leq$ $\omega$ $\leq$ 100 Hz in the previous one. This difference enables a large extension of the scaling regions to the smaller-$\varepsilon$ side: for instance, the scaling region of the dynamic scaling of $\chi''$ at zero field in the present study, $10^{-1}$ $\lesssim$ $\omega\varepsilon^{-z\nu}$ $\lesssim$ $10^{8}$, shown in Fig. \ref{ImAcSus}(c), is largely extended from that in the previous study (0.3 $\lesssim$ $\omega\varepsilon^{-z\nu}$ $\lesssim$ $10^{4}$). As expected from the asymptotic forms of $\chinac{2}$ and $\chi''$, Eqs. (\ref{AcNLSAsymptote}) and (\ref{ImSusAsympt}), the dynamic critical exponent $z\nu$ is strongly affected by the data close to $\Tg$, and thus, the larger $z\nu$, $\sim$ 3.4 in the present study and $\sim$ 2 in the previous one is obtained. The estimation of $\Tg$, and the consequent estimation of the crossover exponent $\phi$, is also sensitive to the data close to $\Tg$, and slightly different results are obtained. On the other hand, the static critical exponents $\gamma$ and $\beta$ are mostly affected by high-$\varepsilon$ (high-$T$) and weakly frequency dependent data, and consequently, almost the same values are obtained in the present and previous studies. 

The influence of the $T$- and $H$-regions on the critical exponents was formerly investigated in the static scaling analyses of the canonical SG Ag(Mn), and  the possibility for the derivation of different critical exponents in high $\varepsilon$ and $H$ ranges from those in low $\varepsilon$ and $H$ ranges was discussed \cite{BouchiatAgMnJPhys}. 
The critical scaling form of physical quantities is an asymptotic feature in the vicinity of the phase-transition temperature, and thus, the present dynamic scaling result using experimental data closer to $\Tg$ is more reliable and genuine. 

\section{Conclusion}

We performed a detailed ac susceptibility measurement of a model magnet of the long-range RKKY Ising SG \dyyrusin{0.103}{0.897} . Dynamic scaling analyses of linear and nonlinear ac susceptibilities in the limited $T$, $H$, and $\omega$ regions, where the critical scaling analyses work appropriately, clearly reveal the existence of the SG phase transition in both zero and finite fields with the same universality class, which indicates the RSB in the long-range RKKY Ising SG. The set of critical exponents (the static critical exponents $\gamma$ $\sim$ 1, $\beta$ $\sim$ 1, and $\delta$ $\sim$ 2, dynamic critical exponent $z\nu$ $\sim$ 3.4, and crossover exponent $\phi$ $\sim$ 2) was obtained. The scaling relation $\phi$ $=$ $\gamma + \beta$ was found within a margin of experimental errors, suggesting that the long-range RKKY Ising SG belongs to a non- or marginal-mean-field universality class.

\section*{Acknowledgment}

This work is partially supported by JSPS KAKENHI Grant-in-Aid for Scientific Research (C) (Grand No. 15K05210). 

\appendix
\section{Dynamic scaling of the nonlinear ac susceptibility}
\label{AppFNLSDS}

The full nonlinear dc susceptibility $\chinldc(T,H)$ $\equiv$ $\Mdc(T,H)/H-\chindc{0}(T)$ corresponds to the SG order parameter under its conjugate field $H^{2}$ and obeys the static scaling \cite{SuzukiPTP,ChalupaSSC}, 
\begin{equation}
\chinldc(T,H) = \varepsilon^{\beta}F(H^{2}\varepsilon^{-\beta\delta}) ,
\label{DcFNLSStaticScaling}
\end{equation}
where $\varepsilon$ $\equiv$ $T/\Tg -1$ is a reduced temperature, $\beta$ and $\delta$ are critical exponents of the order parameter against the temperature and conjugate field at $\Tg$, respectively,  and $F(x)$ is a scaling function. Equation (\ref{DcFNLSStaticScaling}) is expanded as a series of $H^{2}$, 
\begin{equation}
\chinldc(T,H) = \sum_{n > 0} \chindc{2n}(T) H^{2n} \propto \sum_{n > 0} \varepsilon^{\beta(1-n\delta)}H^{2n} ,
\label{DcFNLSExp}
\end{equation}
where the first and higher-order nonlinear susceptibility coefficients $\chindc{2n}(T)$ diverge as $\varepsilon^{\beta(1-n\delta)}$ $=$ $\varepsilon^{-\gamma-(n-1)\beta\delta}$ ($\beta\delta$ $=$ $\gamma + \beta$). The first nonlinear susceptibility coefficient $\chindc{2}$ is a coefficient of the primary term of Eq. (\ref{DcFNLSExp}) and corresponds to the SG order parameter susceptibility diverging  with the critical exponent $\gamma$.  

According to the dynamic scaling hypothesis of second-order phase transitions \cite{HohenDynamicCriticalPhenomena}, the dynamic order parameter susceptibility is described by a scaling function of $\omega\tau$ $=$ $\omega\varepsilon^{-z\nu}$, where $\nu$ is a critical exponent of the correlation length ($\xi$ $\propto$ $\varepsilon^{-\nu}$) and $z$ is a dynamic critical exponent relating the correlation length to the characteristic relaxation time ($\tau$ $\propto$ $\xi^{z}$). In the case of a SG, the divergence of the first nonlinear dc susceptibility coefficient can be extended to a dynamic one as  
\begin{equation}
\chinac{2}(\omega;T) = \varepsilon^{-\gamma} f_{2}(\omega\varepsilon^{-z\nu}) . 
\label{AcNLSDynamicScaling}
\end{equation}
The scaling function $f_{2}(x)$ has asymptotic forms for $x$ $\rightarrow$ 0 ($T$ $\rightarrow$ $\infty$ or $\omega$ $\rightarrow$ 0) and $x$ $\rightarrow$ $\infty$ ($T$ $\rightarrow$ $\Tg$ or $\omega$ $\rightarrow$ $\infty$), 
\begin{equation}
f_{2}(x) \propto \left\{
\begin{array}{ll}
\mbox{const}  & (x \rightarrow 0) , \\
x^{-\gamma/z\nu} & (x \rightarrow \infty ) ,
\end{array}
\right. 
\label{AcNLSDynamicScalingFunc}
\end{equation}
and thus, the first nonlinear ac susceptibility coefficient diverges as 
\begin{equation}
\chinac{2} (\omega;T) \propto \left\{
\begin{array}{llll}
\varepsilon^{-\gamma}  & (\omega \rightarrow 0 & \mbox{or} & T \rightarrow \infty ) , \\
\omega^{-\gamma/z\nu} & (\omega \rightarrow \infty & \mbox{or} & T \rightarrow \Tg ) .
\end{array}
\right. 
\label{AcNLSAsymptote}
\end{equation}
The asymptote for $\omega$ $\rightarrow$ $0$ or $T$ $\rightarrow$ $\infty$ in Eq. (\ref{AcNLSAsymptote}) corresponds to  the divergent behavior of $\chindc{2}$. 

The dynamic scaling form of $\chinac{2}$ can be generalized to arbitrary $n$ ($>$ $0$) as, 
\begin{align}
\chinac{2n} (\omega;T) &= \varepsilon^{-\gamma-(n-1)\beta\delta} f_{2n} (\omega\varepsilon^{-z\nu}) \notag \\
&= \varepsilon^{\beta(1-n\delta)} f_{2n} (\omega\varepsilon^{-z\nu})
\label{AcHOSDynamicScaling}
\end{align}
where $f_{2n}(x)$ is the scaling function and has asymptotic forms, 
\begin{align}
f_{2n}(x) &\propto \left\{
\begin{array}{ll}
\mbox{const} & (x \rightarrow 0) , \\
x^{\beta(1-n\delta)/z\nu} & (x \rightarrow \infty ) .
\end{array}
\right. .
\label{AcHOSDynamicScalingFunc}
\end{align}
Combining Eq. (\ref{AcHOSDynamicScaling}) with the definition of the full nonlinear ac susceptibility $\chinlac(\omega;T,H)$ $\equiv$ $\chiac(\omega;T,H) - \chinac{0}(T)$, the scaling form of $\chinlac(\omega;T,H)$ is obtained as, 
\begin{align}
\chinlac(\omega;T,H) &= \sum_{n > 0} (2n+1) \chinac{2n}(\omega;T)H^{2n} \notag \\
 &= \sum_{n > 0} (2n+1) \varepsilon^{\beta(1-n\delta)} f_{2n}(\omega\varepsilon^{-z\nu}) H^{2n} \notag \\
 &= \varepsilon^{\beta} \tilde{F} (\omega\varepsilon^{-z\nu}, H^{2}\varepsilon^{-\beta\delta}) .
\label{AcFNLSDynamicScaling}
\end{align}  
Consequently, the scaling form of $\chinlac(\omega;T,H)$ at $\Tg$ is deduced using Eqs. (\ref{AcHOSDynamicScalingFunc}) and (\ref{AcFNLSDynamicScaling}) as 
\begin{align}
\chinlac(\omega; \Tg, H) &\rightarrow \sum_{n > 0} (2n+1) \omega^{\beta(1-n\delta)/z\nu}H^{2n} \notag \\
&= \omega^{\beta/z\nu} g(H^{2}\omega^{-\beta\delta/z\nu}) \notag \\
&= H^{2/\delta} h(\omega H^{-2z\nu/\beta\delta}) .
\label{AcFNLSAtTgDynamicScaling}
\end{align}
The scaling function $h(x)$ should asymptote for $x$ $\rightarrow$ $0$ and $\infty$, 
\begin{equation}
h(x) \propto \left\{
\begin{array}{ll}
\mbox{const} & (x \rightarrow 0) , \\
x^{-\gamma/z\nu} & (x \rightarrow \infty) ,
\end{array}
\right. 
\label{AcFNLSAtTgDynamicScalingFunc}
\end{equation}
leading asymptotic forms of $\chinlac(\omega;\Tg, H)$  
\begin{equation}
\chinlac(\omega; \Tg, H) \propto \left\{
\begin{array}{llll}
H^{2/\delta} & (\omega \rightarrow 0 & \mbox{or} & H \rightarrow \infty) , \\
\omega^{-\gamma/z\nu}H^{2} & (\omega \rightarrow \infty & \mbox{or} & H \rightarrow 0) .
\end{array}
\right.
\label{AcFNLSAtTgAsymptote}
\end{equation}
The former form corresponds to the critical field dependence in the static limit, and the latter one corresponds to the zero-field limit as $\chinac{2}(\omega ; T \rightarrow \Tg)H^{2}$.   

\vspace*{18pt}
\section{Dynamic scaling of the imaginary part of the ac susceptibility}
\label{AppIACSUSDS}

The spin autocorrelation function $q$ is a dynamic SG order parameter and obeys the dynamic scaling by $t/ \tau$ as \cite{HohenDynamicCriticalPhenomena,OgielskiSGDScalingPRB} 
\begin{align}
q(t;T) &= t^{-\beta/z\nu} Q(t/\tau) .
\label{AutoSpinCorDynamicScaling}
\end{align}
Using linear-response theory, the imaginary part of the ac susceptibility $\chi''$ and its scaling form are derived from $q$ as 
\begin{align}
\chi''(\omega;T) &= \chieq(T)\omega \int_{0}^{\infty} dt q(t;T) \cos\omega t \notag \\
&= \chieq(T)\omega\tau^{1-\beta/z\nu} \int_{0}^{\infty} dyy^{-\beta/z\nu}Q(y)\cos\omega\tau y \notag \\
&= \chieq(T)\tau^{-\beta/z\nu} K(\omega\tau) \notag \\
&= \chieq(T)\varepsilon^{\beta} K(\omega\varepsilon^{-z\nu}) , 
\label{ImSusDynamicScaling}
\end{align}
where the scaling function is given by 
\begin{align}
K(x)  &= x\int_{0}^{\infty} dy y^{-\beta/z\nu}Q(y)\cos xy .
\label{ImSusDynamicScalingFunc}
\end{align}
The spin autocorrelation function $q(t)$ is expected to decay in power-law fashion as $t^{-\beta/z\nu}$  at $T$ $=$ $\Tg$. And $\chi''$ is proportional to $\omega$ far from $\Tg$. Thus, the scaling function $K(x)$ should asymptote as 
\begin{align} 
K(x) &\propto \left\{ 
\begin{array}{ll}
x & (x \rightarrow 0) . \\
x^{\beta/z\nu} & (x \rightarrow \infty) ,
\end{array}
\right.  
\label{ImSusDynamicScalingFuncAsympt}
\end{align}
which leads to the following asymptotic forms of $\chi''/\chieq$: 
\begin{align}
\frac{\chi''(\omega;T)}{\chieq(T)} &\propto \left\{
\begin{array}{llll}
\omega\varepsilon^{\beta-z\nu} & (\omega \rightarrow 0 & \mbox{or} & T \rightarrow \infty ) , \\
\omega^{\beta/z\nu} & (\omega \rightarrow \infty & \mbox{or} & T \rightarrow \Tg ) .
\end{array}
\right. 
\label{ImSusAsympt}
\end{align}

If the SG phase transition under a finite magnetic field exists, the characteristic relaxation time at $T$ and $H$, $\tau(T,H)$, shows  a critical divergent behavior to a certain phase-transition temperature in the field $\Tg(H)$ as $\tau(T,H)$ $\propto$ $\varepsilon(H)^{-z\nu}$, where $\varepsilon(H)$ $\equiv$ $[T/\Tg(H) -1]$. Thus, the dynamic scaling form (\ref{ImSusDynamicScaling}) can be rewritten as 
\begin{align}
\frac{\chi''(\omega;T,H)}{\chieq(T,H)} &= \varepsilon(H)^{\beta} K\left[\omega\varepsilon(H)^{-z\nu}\right] . 
\label{ImSusDynamicScalingEpsilon}
\end{align}


\bibliography{DYRS_PRB_manuscript02}

\end{document}